\documentclass[11pt,showpacs,preprintnumbers,amsmath,amssymb,prd,nofootinbib,superscriptaddress]{revtex4-2}

\usepackage{dcolumn}
\usepackage{bm}
\usepackage{ifpdf}
\usepackage{hyperref}
\usepackage{bm}
\usepackage{xcolor,color,graphicx,graphics}
\usepackage[spanish,english]{babel}
\usepackage[latin1]{inputenc}
\usepackage[OT1]{fontenc}
\usepackage{latexsym,amssymb,amsmath,amsfonts}
\usepackage{makeidx}
\usepackage{epsfig,subfigure}
\usepackage{natbib}
\usepackage{epstopdf}
\usepackage{mathrsfs}
\usepackage{hyperref}
\hypersetup{colorlinks=true, linkcolor=blue, citecolor=blue, urlcolor=blue}
\usepackage{enumerate}

\everymath{\displaystyle}
\usepackage{graphicx}

\usepackage[T1]{fontenc}
\usepackage{amsmath}
\usepackage{amssymb}
\usepackage{graphicx}
\usepackage{xcolor}

\newcommand{\bea}{\begin{eqnarray}}
\newcommand{\eea}{\end{eqnarray}}

\newcommand{\orcid}[1]{\href{https://orcid.org/#1}{\includegraphics[width=10pt]{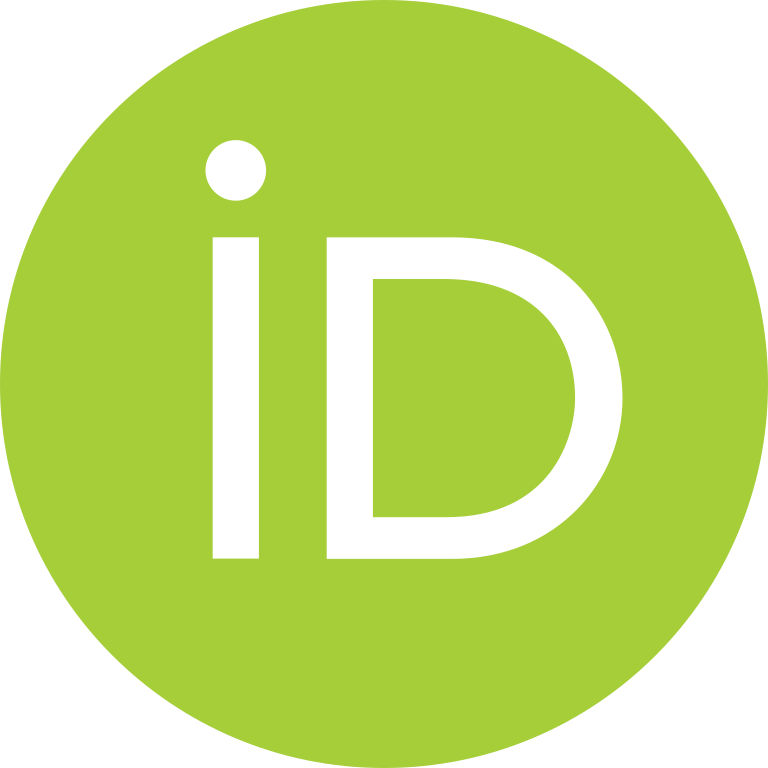}}}

\usepackage{fixmath}

\newcommand{\pr}[1]{\ensuremath{\left[#1\right]}}
\newcommand{\pc}[1]{\ensuremath{\left(#1\right)}}

\newcommand{\R}{R_{\mu\nu}}
\newcommand{\T}{T^{\mu\nu}}

\begin{document}

\title{G\"{o}del-type solutions in $f(R,T,R_{\mu\nu} T^{\mu\nu}$) gravity}

\author{J. S. Gon\c{c}alves \orcid{0000-0001-6704-2748}}
\email{junior@fisica.ufmt.br}
\affiliation{Instituto de F\'{\i}sica, Universidade Federal de Mato Grosso,\\
78060-900, Cuiab\'{a}, Mato Grosso, Brazil}

\author{A. F. Santos \orcid{0000-0002-2505-5273}}
\email{alesandroferreira@fisica.ufmt.br}
\affiliation{Instituto de F\'{\i}sica, Universidade Federal de Mato Grosso,\\
78060-900, Cuiab\'{a}, Mato Grosso, Brazil}

\begin{abstract}

In this paper, $f(R,T,R_{\mu\nu} T^{\mu\nu}$) gravity is considered. It is a modified theory of gravity that exhibits a strong coupling of gravitational and matter fields. Therefore, if gravity is governed by this model a number of issues must be re-examined. In this context, the question of causality and its violation is studied. Such analysis is carried out using the G\"{o}del-type solutions. It is shown that this model allows both causal and non-causal solutions. These solutions depend directly on the content of matter present in the universe. For the non-causal solution, a critical radius is calculated, beyond which causality is violated. Taking different matter contents, an infinite critical radius emerges that leads to a causal solution. In this causal solution, a natural relationship emerges between the parameters that determine the matter considered.

\end{abstract}

\maketitle

\section{Introduction}

A widely accepted fact in the scientific community is the accelerated expansion of the universe which is strongly supported by observations \cite{Riess,Per,Adam,Cole,Ande,expansao_ace,expansao_ace2}. Since the General Relativity theory (GR) does not adequately explain this phenomenon, two ways have been investigated: (i) exotic component of the matter, called dark energy, has been  considered, or (ii) alternative models to the GR have been proposed. In this paper, the study developed considers the second case, i.e. modified gravity theory. The $f(R)$ theory is the simplest and most popular way to modify GR \cite{faraoni1}. Several studies have already been done on this type of theory, such as the Newtonian limit \cite{newtonian_limit}, gravitational stability \cite{gravitational_stability}, cosmological evolution of solar-system tests \cite{solar_sistem}, inflation \cite{inflation_FR}, among others. Another theory is $f(R,T)$ gravity \cite{PhysRevD.84.024020}, which is a generalization of $f(R)$ theory. In this case, the Ricci scalar $R$ in the Hilbert-Einstein action is changed by some function dependent on the Ricci scalar $R$ and the trace of the energy-momentum tensor $T$. The dependence on $T$ can be induced by imperfect fluids or conformal anomalies arising from quantum effects \cite{quantum_cosmology,energy_conditions}. This gravitational theory has also been considered in the context of Palatini formalism. In general, Palatini-type theories are very attractive as they present a good analysis of the initial and current dynamics of the universe \cite{palatini_motivation1,fR_palatini_motivation2,primeiro_frt}.

In the present work, an extension of $f(R,T)$ gravity is proposed. The action is described by a function $f(R,T,\R\T)$ where $\R$ and $\T$ are the Ricci tensor and the energy-momentum tensor, respectively. The stability of this theory has been investigated in various contexts, such as, stability analysis of stellar radiating filaments \cite{frtRT_motivation1}, stability of cylindrical stellar model \cite{frtRT_motivation2} and stability of Einstein Universe \cite{frtRT_motivation3}. Furthermore, $f(R,T,\R\T)$ gravity has already been investigated in $\Lambda$CDM Universe \cite{frtRT_motivation4}, cosmic evolution in the background of non-minimal coupling \cite{frtRT_motivation5}, among others. However, study about causality and its violation has not been investigated in this theory. This is an important test to be performed by all alternative gravitational theories, since GR allows for such a discussion.

To investigate the causality problem in $f(R,T,\R\T)$ gravity the G\"{o}del-type metric \cite{tipo_godel1} is considered. It is a generalization of the G\"{o}del metric proposed by Kurt G\"{o}del, in 1949 \cite{godel}. It is the first exact solution for the GR with rotating matter. This metric leads to the possibility of Closed Timelike Curves (CTCs). These CTCs are not exclusive to the G\"{o}del solution, they appear in other cosmological models that have some hyperbolic or spherical symmetry, such as Kerr black hole, Van-Stockum model, cosmic string, among others \cite{ctc1,ctc2}. The G\"{o}del-type metric provides more information about the violation of causality. It allows the calculation of a critical radius $r_c$ that defines causal and non-causal regions. The causality problem has already been studied in several different models. In  \cite{fr_and_godel} it shows that G\"{o}del and type-G\"{o}del are solutions to the $f(R)$ theory, as well as in  \cite{kessence_and_godel} that both G\"{o}del and type-G\"{o}del are solutions to the $k$-essence theory. Causality is also discussed in Chern-Simons gravity \cite{chersimon_and_godel1, chersimon_and_godel2}, $f(T)$ gravity \cite{ft_and_godel}, $f(R,T)$ gravity \cite{frt_and_godel}, bumblebee gravity \cite{bumblebeee_and_godel}, Horava-Lifshitz gravity \cite{horava_and_godel}, Brans-Dicke theory
\cite{brans_and_godel} and $f(R,Q)$ gravity \cite{frq_and_godel}. More recently causality is discuss in $f(R,\phi,X)$ and $f(R,T)$ Palatini gravity \cite{frt_palatini,fr_phi_X}. In this paper, the main objective is to investigate whether the $f(R,T,\R\T)$ gravity allows for causality violation. The analysis is divided into two parts, first the content of matter is just a perfect fluid and then the content of matter is a perfect fluid plus a scalar field. 

The present paper is organized as follows. In Section II, $f(R,T,\R\T)$ gravity is introduced and the field equations are derived. In Section III, the G\"{o}del-type metric is discussed. Considering a perfect fluid as matter content, the standard G\"{o}del solution is obtained. In Section IV, the problem of causality is verified for a perfect fluid plus a scalar field as the content of matter. This content of matter leads to a causal G\"{o}del-type solution. In Section V, remarks and conclusions are presented.

\section{$f(R,T,\R\T)$ Gravity}

In this section, the gravitational field equations for $f(R,T,\R\T)$ gravity are obtained. The action that describes this theory is
\begin{equation}\label{action_1}
    S=\frac{1}{2\kappa^2}\int d^4x \sqrt{-g} \pr{f(R,T,\R\T) + \mathcal{L}_m},
\end{equation}
where $\kappa^2 = 8\pi G$, $g$ is the determinant of the metric tensor $g_{\mu\nu}$, $R$ is the Ricci scalar, $f(R,T,\R\T)$ is a function that depends on the Ricci scalar, the trace of energy-momentum and the tensor product $\R\T$, respectively, and $\mathcal{L}_m$ is the matter Lagrangian.

Varying the action (\ref{action_1}) in relation to the metric $g^{\mu\nu}$ leads to
\begin{equation}\label{field_1}
     \delta S = \frac{1}{2\kappa^2} \int d^4x \pr{\delta \sqrt{-g} f + \sqrt{-g} \pc{f_R \delta R + f_T\delta T + f_Q \delta \R \T }} + \int d^4x \delta \pc{\sqrt{-g} \mathcal{L}_m} , 
\end{equation}
with $f_R \equiv \frac{\partial f}{\partial R}$, $f_T \equiv \frac{\partial f}{\partial T}$ and $f_Q \equiv \frac{\partial f}{\partial Q}$, where $Q \equiv \R \T$. 

Considering the  definition of the energy-momentum tensor, we get
\begin{equation}\label{tensor energia}
    T_{\mu\nu} = \frac{-2}{\sqrt{-g}} \frac{\partial \pc{\sqrt{-g}\mathcal{L}_m}}{\partial g^{\mu\nu}} = -2 \frac{\partial \mathcal{L}_m}{\partial g^{\mu\nu}} + g_{\mu\nu} \mathcal{L}_m,
\end{equation}
where $\mathcal{L}_m$ is assumed to be dependent only on the metric and not on the first derivatives. The variation in the energy-momentum tensor trace is given as  
\begin{align}
\delta T &= \delta \pc{g^{\alpha \beta} T_{\alpha\beta}},\\
&= \pc{\frac{\delta g^{\alpha\beta}}{\delta g^{\mu\nu}}} T_{\alpha\beta} + \frac{\delta T_{\alpha\beta}}{\delta g^{\mu\nu}} g^{\alpha\beta},\\
&=\pc{T_{\mu\nu} + \Theta_{\mu\nu}} ,
\end{align}
with  $\Theta_{\mu\nu}$ being a tensor defined as
\begin{equation}
    \Theta_{\mu\nu} \equiv \frac{\delta T_{\alpha\beta}}{\delta g^{\mu\nu}} g^{\alpha\beta}.
\end{equation}
Using Eq. (\ref{tensor energia}) this tensor can be written as
\begin{equation}
    \Theta_{\mu\nu} = -2T_{\mu\nu} + g_{\mu\nu} \mathcal{L}_m  -2 g^{\alpha\beta} \frac{\partial^2 \mathcal{L}_m}{\partial g^{\mu\nu}\partial g^{\alpha\beta}}.
\end{equation}

The variations $\delta R$ and $\delta T$ are known \cite{reviewbook}. They are given as
\begin{equation}
    f_R \delta R + f_T \delta T = \pr{\R f_R + \pc{g_{\mu\nu} \Box - \nabla_ \mu \nabla_ \nu}f_R + \pc{T_{\mu\nu} + \Theta_{\mu\nu}}f_T} \delta g^{\mu\nu}.
\end{equation}
The variation corresponding to the term $\R\T$ is composed of two parts, i.e.
\begin{equation}
    f_Q \T \delta \R = \pr{\frac{1}{2} \Box (f_Q T_{\mu\nu}) + g_{\mu\nu} \nabla_\alpha \nabla_\beta (f_Q T^{\alpha \beta}) - \nabla_\alpha \nabla_\nu (f_Q T^\alpha_{\mu})}
\end{equation}
and
\begin{equation}
    f_Q \R \delta \T = f_Q \pr{-G_{\mu\nu} \mathcal{L}_m - \frac{1}{2} RT_{\mu\nu} + 2R^\alpha_{\mu} T_{\alpha \nu} - 2R^{\alpha \beta} \frac{\delta^2 \mathcal{L}_m}{\delta g^{\mu\nu}\delta g^{\alpha \beta}}} \delta g^{\mu\nu}.
\end{equation}

Considering the perfect fluid as matter content, it is assumed that the Lagrangian of the matter is $\mathcal{L}_m = -p$. This choice implies that
\begin{equation}
    \frac{\delta^2 \mathcal{L}_m}{\delta g^{\mu\nu}\delta g^{\alpha \beta}} =0.
\end{equation}
Then the tensor $\Theta_{\mu\nu}$ becomes
\begin{equation}
    \Theta_{\mu\nu} = -2T_{\mu\nu} - pg_{\mu\nu}.\label{theta}
\end{equation}

Taking these results to Eq. (\ref{field_1}), the complete set of the field equations is given by
\begin{eqnarray}
 &&R_{\mu\nu} f_R -\frac{1}{2} g_{\mu\nu} f + \pc{g_{\mu\nu} \Box - \nabla_\mu \nabla_\nu} f_R + f_T\pc{T_{\mu\nu} + \Theta_{\mu\nu}}\nonumber\\
 && +\frac{1}{2} \pc{\Box (T_{\mu\nu} f_Q) + g_{\mu\nu} \nabla_\alpha \nabla_\beta (T^{\alpha \beta} f_Q) } - \nabla_\alpha (\nabla_{(\mu} T^{\alpha}_{\nu )}f_Q) + \Xi_{\mu\nu} f_Q = \kappa^2 T_{\mu\nu},\label{FE}
\end{eqnarray}
with
\begin{equation}
        \Xi_{\mu\nu} = -G_{\mu\nu} p -\frac{1}{2}RT_{\mu\nu} + 2R^{\alpha}_{(\mu}T_{\nu )\alpha}.\label{RT}
\end{equation}

In the next section, these field equations will be studied considering the cosmological background described by the G\"{o}del-type solutions.

\section{G\"{o}del-type Metric and Perfect Fluid}

Here the modified Einstein equations are solved considering the G\"{o}del-type metric and a perfect fluid as the content of matter. Before solving the field equations obtained in the previous section, the main characteristics of this solution are presented.

The G\"{o}del-type solution is described by the line element
\begin{equation}\label{type_godel}
    ds^2 = -dt^2 - 2H(r) dtd\phi + dr^2 + \left(D^2(r)-H^2(r)\right) d\phi^2 + dz^2,
\end{equation}
where $H(r)$ and $D(r)$ are functions that satisfies the relations
\bea
\frac{H'(r)}{D(r)}=2\omega \quad \mathrm{and}\quad \frac{D''(r)}{D(r)}=m^2.
\eea
The prime implies the derivative with respect to $r$. It is important to note that $\omega$ and $m^2$ are free parameters that characterize all properties of the metric \cite{tipo_godel1,godel_fRT}. This is a generalization of the solution proposed by G\"{o}del. The G\"{o}del metric is an exact solution of Einstein equations for a homogeneous rotating universe. This solution leads to the possibility of Closed Time-like Curves (CTCs) that allow violation of causality. As an immediate consequence,  this makes time travel theoretically possible in this space-time. For a more detailed investigation of causality, as well as its violation, the G\"{o}del-type solution is used. From the parameters $\omega$ and $m^2$ three different classes are defined, i.e. hyperbolic, trigonometric and linear class \cite{tipo_godel1}. Here only the hyperbolic class is considered. For this class, the functions $H(r)$ and $D(r)$ are given as
\begin{align}
    H(r) &= \frac{4\omega}{m^2} senh^2 \left(\frac{mr}{2} \right),\label{condition1} \\
     D(r) &= \frac{1}{m} senh(mr)\label{condition2}.
\end{align}

An important quantity that is calculated from G\"{o}del-type solution is the critical radius, beyond which CTCs exist. It is defined as
\begin{equation}
    r_c = \frac{2}{m} senh ^{-1} \left(\frac{4\omega^2}{m^2}-1\right)^{-1}.
\end{equation}
From this, two relationships between the parameters $\omega^2$ and $m^2$ are interesting. (i) $m^2=2\omega^2$ that leads to the G\"{o}del solution with a finite critical radius given by
\begin{equation}
    r_c = \frac{2}{m} senh ^{-1} \left(1\right).
\end{equation}
As a consequence, non-causal regions are allowed. And (ii) $m^2=4\omega^2$ implies an infinite critical radius. In other words, this condition leads to a causal solution.

For simplicity, a new basis (a local Lorentz co-frame) is chosen, such that the metric becomes  \cite{tipo_godel1},
\begin{equation}
    ds^2 = \eta_{AB} \theta^A \theta^B = (\theta^{(0)})^2 - (\theta^{(1)})^2 - (\theta^{(2)})^2 - (\theta^{(3)})^2, \label{frame}
\end{equation}
where $\eta_{AB}$ is the Minkowski metric and $ \theta^A = e^A\ _{\mu}\ dx^\mu$ with
\begin{align}
    \theta^{(0)} &= dt + H(r)d\phi,\nonumber \\ \theta^{(1)} &= dr, \nonumber\\ \theta^{(2)} &= D(r)d\phi,\nonumber \\ \theta^{(3)} &= dz,    \label{coframe}
\end{align}
and $e^A\ _{\mu}$ is the tetrad that satisfies the relation  $ e^A\,_\mu e^\mu\,_B=\delta^A_B$. Here the capital Latin letters label Lorentz indices and run from 0 to 3.

In the local Lorentz co-frame (\ref{coframe}), the non-zero Ricci tensor components are
\begin{eqnarray}
     R_{(0)(0)} &=& \frac{1}{2} \pc{\frac{H'}{D}}^2, \nonumber\\ 
    R_{(1)(1)} &=& R_{(2)(2)} = R_{(0)(0)} - \frac{D''}{D}.
\end{eqnarray}
It is interesting to note that all components are constants, i.e.
\begin{equation}
    R_{(0)(0)} = 2\omega^2, \ R_{(1)(1)} = \ R_{(2)(2)}= 2\omega^2 - m^2.
\end{equation}

The non-vanishing components of the Einstein tensor in flat (local) space-time take the form,
\begin{align}
     G_{(0)(0)} &= 3\omega^2-m^2, \\
    G_{(1)(1)} &= \omega^2,\\
    G_{(2)(2)} &= \omega^2, \\
    G_{(3)(3)} &= m^2-\omega^2,
\end{align}
where $G_{AB}=e^\mu_A e^\nu_B G_{\mu\nu}$ has been used. The Ricci scalar is $R = 2(m^2 - w^2)$.

Assuming that the content of matter is a perfect fluid, the energy-momentum tensor that describes it is
\begin{eqnarray}
    T_{AB}=(\rho+p)u_A u_B-p\eta_{AB},\label{EMT}
    \end{eqnarray}
with $u_A = (1,0,0,0)$ being the four-velocity of fluid. Thus, in flat space-time the tensor given in Eq. (\ref{theta}) becomes
\begin{eqnarray}
\Theta_{AB} = -2T_{AB} - p\eta_{AB}.\label{Th}
\end{eqnarray}
Taking the Minkowski metric as
\begin{equation}
    \eta_{AB} = \left(
    \begin{array}{cccc}
    1   & 0    & 0  & 0 \\
    0  & -1 & 0 & 0  \\
    0   & 0 & -1 & 0 \\
    0 & 0 & 0 & -1
    \end{array}\right),
\end{equation}
the components of the tensors (\ref{EMT}) and (\ref{Th}) are
\begin{equation}
    T_{AB} =\left(
    \begin{array}{cccc}
    \rho   & 0    & 0  & 0 \\
    0  & p & 0 & 0  \\
    0   & 0 & p & 0 \\
    0 & 0 & 0 & p
    \end{array}\right) ;\ \  
    \Theta_{AB} = \left(
    \begin{array}{cccc}
    -2\rho - p   & 0    & 0  & 0 \\
    0  & -p & 0 & 0  \\
    0   & 0 & -p & 0 \\
    0 & 0 & 0 & -p
    \end{array}\right).\label{33}
\end{equation}
The trace of $T_{AB}$ and $\Theta_{AB}$ are $T= \rho -3p$ and $\Theta = 2(p-\rho)$. Using the components of the energy-momentum tensor and the Ricci tensor, the components of $Q=R_{AB}T^{AB}$ are given as
\begin{equation}
    Q =  \left(\begin{matrix}2 \omega^{2} \rho & 0 & 0 & 0\\0 &  \left(2 \omega^{2}- m^{2} \right)p & 0 & 0\\0 & 0 &   \left(2 \omega^{2}- m^{2} \right)p & 0\\0 & 0 & 0 & 0\end{matrix}\right).
\end{equation}
It should be noted that the derivatives of the functions $f_R$ and $f_Q$ vanish since the Ricci scalar and $Q$ are constants. Then the field equations (\ref{FE}) are rewritten as
\begin{align}\label{field eq type_godel}
    R_{AB} f_R = \kappa^2 T_{AB} + \frac{1}{2} \eta_{AB} f - f_T\pc{T_{AB} + \Theta_{AB}} - \Xi_{AB} f_Q.
\end{align}
Taking the trace of this equation leads to
\begin{equation}\label{trace}
    R f_R = \kappa^2 T +2 f- f_T\pc{T + \Theta} - \Xi f_Q,
\end{equation}
where $\Xi = \eta^{AB}  \Xi_{AB}$. Combining these equations we have
\begin{equation}\label{eq_campo_completa}
    f_R G_{AB} = \kappa^2 T_{AB} - f_T\pc{T_{AB} + \Theta_{AB}} - \Xi_{AB} f_Q -\frac{1}{2} \eta_{AB} \pr{f + \kappa^2 T - f_T (T + \Theta) -\Xi f_Q}
\end{equation}
with $G_{AB} = R_{AB} - \frac{1}{2} R \eta_{AB}$ being the Einstein tensor. In order to solve the set of field equations, let us explicitly write Eq. (\ref{RT}) in flat space-time. Then
\begin{equation}
    \Xi_{AB} = \left(\begin{matrix}-( 3\omega^{2}-m^{2})p - (m^{2} - 5 \omega^{2}) \rho & 0 & 0 & 0\\0 &  \left( m^{2} -4 \omega^{2}\right)p & 0 & 0\\0 & 0 &  \left(m^{2} -4 \omega^{2}\right)p & 0\\0 & 0 & 0 & -2\left(m^2-\omega^2\right)p\end{matrix}\right).
\end{equation}
As a consequence $\Xi=\left(m^2+3\omega^2\right)p-\left(m^2-5\omega^2\right)\rho$.

Using these results the field equations (\ref{eq_campo_completa}) become
\bea
2f_R\left(3\omega^2-m^2\right)+f&=&\kappa^2\left(\rho+3p\right)+f_T\left(\rho+p\right)+f_Q\left[\left(9\omega^2-m^2\right)p+\left(m^2-5\omega^2\right)\rho\right],\label{1}\\
2f_R\omega^2-f &=&\kappa^2\left(\rho-p\right)+f_T\left(\rho+p\right)+f_Q\left[\left(5\omega^2-3m^2\right)p+\left(m^2-5\omega^2\right)\rho\right],\label{2}\\
2f_R\left(m^2-\omega^2\right)-f&=&\kappa^2\left(\rho-p\right)+f_T\left(\rho+p\right)+f_Q\left[\left(3m^2-7\omega^2\right)p+\left(m^2-5\omega^2\right)\rho\right].\label{3}
\eea
Equations (\ref{2}) and (\ref{3}) lead to
\bea
\left(2\omega^2-m^2\right)\left(2f_R-6pf_Q\right)=0.\label{4}
\eea
Assuming $f_R>0$ and $f_Q>0$, Eq. (\ref{4}) gives us
\begin{equation}
    m^2=2\omega^2.
\end{equation}
This condition defines the G\"{o}del solution, and the remaining field equations read
\bea
m^2f_R+f&=&\kappa^2\left(\rho+3p\right)+f_T\left(\rho+p\right)+f_Q\left(7p-3\rho\right)\frac{m^2}{2}, \label{5}\\
m^2f_R-f&=&\kappa^2\left(\rho-p\right)+f_T\left(\rho+p\right)-f_Q\left(3p+9\rho\right)\frac{m^2}{8}.
\eea
Since these equations allow the G\"{o}del solution, a critical radius, which defines regions where causality is violated, can be calculated. Then the critical radius in $f(R,T, \R\T)$ gravity is 
\bea
r_c=2\sinh^{-1}(1)\sqrt{\frac{2f_R-\frac{f_Q}{16}\left(19p-39\rho\right)}{2\kappa^2\rho+f+2f_T(\rho+p)}}.
\eea
Note that the $r_c$ depends on the gravity theory and the content of matter. It is important to emphasize that this quantity is obtained for any $f(R,T, \R\T)$ gravity. A similar result is obtained for other gravity theories such as $f(R)$ and $f(R,T)$.

It is shown that for a perfect fluid as the content of matter non-causal G\"{o}del curves are unavoidable. From this result, a natural question arises: is there any condition that leads to a causal G\"{o}del-type solution in this gravitational theory? In the next section, such an investigation is developed.

\section{Matter content: Perfect fluid and scalar field}

Here the content of matter is a combination of perfect fluid and scalar field \cite{fr_hybrid,tipo_godel_frt}. The main idea is to investigate the possibility of finding a causal solution in this theory. The total energy-momentum tensor that describes this combination is given as
\bea
 T_{AB} = T_{AB}^M + T_{AB}^S,
\eea
where 
$T_{AB}^M$ and $T_{AB}^S$ are the energy-momentum tensor of the perfect fluid and scalar field, respectively. The Lagrangian for the scalar field $\Phi$ is
\begin{equation}
    \mathcal{L}^S = \eta^{AB} \nabla_A \Phi \nabla_B \Phi, 
\end{equation}
and the corresponding energy-momentum tensor reads
\bea 
T_{AB}^S=\nabla_A \phi \nabla_B \phi - \frac{1}{2} \eta_{AB}\eta^{CD}\nabla_C \phi \nabla_D \phi.
\eea
The energy-momentum tensor for the perfect fluid is given in Eq. (\ref{EMT}). Then the total energy-momentum tensor is
\bea 
T_{AB} = \pc{\rho + p} u_Au_B -p\eta_{AB} + \nabla_A \phi \nabla_B \phi - \frac{1}{2} \eta_{AB}\eta^{CD}\nabla_C \phi \nabla_D \phi.
\eea

The components of the energy-momentum tensor of matter are given in Eq. (\ref{33}), while to write the components associated with the scalar field it is considered 
$\Phi = \epsilon z + \epsilon $, with $\epsilon = const$. Then
\begin{equation}
    T_{00}^S = -T_{11}^S = -T_{22}^S = T_{33}^S = \frac{\epsilon^2}{2}.
\end{equation}
The trace of $T_{AB}$ is given by
\begin{equation}
    T=\rho -3p+\epsilon^2.
\end{equation}

In the same way, the tensor $\Theta_{AB}$ must be rewritten considering the contributions of the scalar field. Thus 
\begin{equation}
    \Theta_{AB} = \Theta_{AB}^M + \Theta_{AB}^S,
\end{equation}
where $\Theta_{AB}^M$ is already known in Eq. (\ref{33}) and $\Theta_{AB}^S$ is given as
\begin{equation}
    \Theta_{AB}^S = - T_{AB}^S + \frac{1}{2} T^S \eta_{AB}, 
\end{equation}
with $T^S$ being the trace of $T_{AB}^S$.

Using the total Lagrangian
\begin{equation}
    \mathcal{L}_m = \mathcal{L}^{PF} + \mathcal{L}^{S} = p - \epsilon^2,
\end{equation}
where $ \mathcal{L}^{PF}$ and $\mathcal{L}^{S}$ are the Lagrangians of the perfect fluid and the scalar field, respectively, the tensor $\Xi_{AB}$ becomes
\begin{equation}
    \Xi_{AB} = -G_{AB} p +G_{AB}\epsilon^2 -\frac{1}{2}RT_{AB} + \eta^{CD} R_{AD} T_{BC} + \eta^{CD} R_{BD} T_{AC}.
\end{equation}

With these results, the field equations (\ref{eq_campo_completa}) are written as
\begin{align}
    f_R G_{AB} &= \kappa^2 \pr {\pc{\rho  + p}u_Au_B -p\eta_{AB} + T_{AB}^S } - \Xi_{AB} f_Q\\
    &-\frac{1}{2} \pr{\kappa^2 \pc{\rho - 3p + \epsilon^2 } +f +f_T \pc{\rho + p - 2\epsilon^2 } - \Xi f_Q }\eta_{AB}\\
    & +f_T \pr{\pc{\rho + p }u_Au_B - \frac{1}{2} \epsilon^2 \eta_{AB}}.
\end{align}
This equation leads to the following set of field equations
\bea
2f_R\left(3\omega^2-m^2\right)+f&=&\kappa^2\left(\rho+3p\right)+f_T\left(\rho+p+\epsilon^2\right)\nonumber\\
&&+f_Q\left[\left(9\omega^2-m^2\right)p+\left(m^2-5\omega^2\right)\rho-2\omega^2\epsilon^2\right],\\
2f_R\omega^2-f&=&\kappa^2\left(\rho-p\right)+f_T\left(\rho+p-\epsilon^2\right)\nonumber\\
&&+f_Q\left[\left(-3m^2+5\omega^2\right)p+\left(m^2-5\omega^2\right)\rho+\left(m^2-2\omega^2\right)\epsilon^2\right],\\
2\left(m^2-\omega^2\right)f_R-f&=&\kappa^2\left(\rho-p+2\epsilon^2\right)+f_T\left(\rho+p-\epsilon^2\right)\nonumber\\
&&+f_Q\left[\left(3m^2-7\omega^2\right)p+\left(m^2-5\omega^2\right)\rho\right].
\eea
For simplicity, after some manipulation, these equations are rewritten as
\bea
\kappa^2\epsilon^2&=&\left(m^2-2\omega^2\right)f_R+f_Q\left[-3p\left(m^2-\omega^2\right)+\frac{\epsilon^2}{2}\left(m^2-2\omega^2\right)\right],\\
\kappa^2p+\frac{1}{2}f_T\epsilon^2&=&\frac{1}{2}\left(2\omega^2-m^2\right)f_R+\frac{1}{2}f+\frac{f_Q}{4}\left[m^2\epsilon^2-\left(4\omega^2+2m^2\right)p\right],\\
\kappa^2\rho+f_T\left(\rho+p-\frac{\epsilon^2}{2}\right)&=&\frac{1}{2}\left(6\omega^2-m^2\right)f_R-\frac{1}{2}f+f_Q\left[2\left(5\omega^2-2m^2\right)p-\frac{\epsilon^2}{2}\left(m^2+2\omega^2\right)\right].
\eea
Assuming that these equations satisfy the conditions $f_R > 0$, $f_T > 0$ and $f_Q > 0$ implies
that the set of equations satisfy a causal solution, i.e.
\begin{equation}
    m^2= 4\omega^2.\label{causal}
\end{equation}
As a consequence of Eq. (\ref{causal}) the critical radius goes to infinity, i.e. $r_c \rightarrow \infty$. Therefore, for this combination of perfect fluid and scalar field as matter content, a causal G\"{o}del-type solution is allowed in $f(R,T,R_{\mu\nu}T^{\mu\nu})$ gravity. In addition, due to the consistency of the equations, a relationship between the constant $\epsilon$ associated with the scalar field and the pressure $p$ of the perfect fluid is required.

It is important to say that by making $f_Q$ go to zero in the field equations, the results already obtained for the $f(R,T)$ theory are recovered. This reinforces the principle of correspondence between theories. 

\section{Discussions and Final Remarks}

In this paper, $f(R,T,\R\T)$ gravity is considered and the causality violation is investigated using the G\"{o}del-type metric. It was shown that this theory generalizes GR implying a geometry-matter coupling. An important question that must be answered by all modified theories is: does the new theory contain all solutions of the GR? In order to answer this question, this work investigated whether an exact solution of GR, such as G\"{o}del metric or G\"{o}del-type metric, holds in $f(R,T,\R\T)$ gravity.

Considering the hyperbolic class of G\"{o}del-type solution and the perfect fluid as matter content, the non-causal G\"{o}del solution is obtained. From this, a  finite critical radius is calculated that leads to the possible existence of CTCs. In this case, the critical radius is a function that depends on the function $f(R,T,\R\T)$ and the derivatives of its variables, $f_R$, $f_T$ and $f_Q$. To investigate causal solutions in this gravitational theory, different content of matter is considered, i.e., a combination between perfect fluid and scalar field. With these considerations, a condition emerges from the set of field equations that implies an infinite critical radius. Thus, the violation of causality is avoided in this theory. Furthermore, the result presented here is a generalization of the results obtained for $f(R)$ and $f(R,T)$ gravities. However, a different condition arises for the causal solution in  $f(R,T,\R\T)$ gravity, namely, there is a certain connection between the contents of matter, i.e., the perfect fluid pressure and the scalar field are related.

\section*{Acknowledgments}

This work by A. F. S. is partially supported by National Council for Scientific and Technological Develo\-pment - CNPq project No. 313400/2020-2. J. S. Gon\c{c}alves thanks CAPES for financial support. 


\global\long\def\link#1#2{\href{http://eudml.org/#1}{#2}}
 \global\long\def\doi#1#2{\href{http://dx.doi.org/#1}{#2}}
 \global\long\def\arXiv#1#2{\href{http://arxiv.org/abs/#1}{arXiv:#1 [#2]}}
 \global\long\def\arXivOld#1{\href{http://arxiv.org/abs/#1}{arXiv:#1}}



\begin{thebibliography}{99}

\bibitem{Riess} A. G. Riess et al., ``Observational Evidence from Supernovae for an Accelerating Universe and a Cosmological Constant,''
\doi{10.1086/300499} {Astron. J. {\bf 116}, 1009 (1998).}

\bibitem{Per} S. Perlmutter et al., ``Discovery of a supernova explosion at half the age of the Universe.''
\doi{10.1038/34124} {Nature {\bf 391}, 51 (1998).}

\bibitem{Adam} R. Adam et al., ``Planck 2015 results,''
\doi{10.1051/0004-6361/201527101} {Astron. Astrophys. {\bf 594}, A1 (2016).}

\bibitem{Cole} S. Cole et al., ``The 2dF Galaxy Redshift Survey: power-spectrum analysis of the final data set and cosmological implications,''
\doi{10.1111/j.1365-2966.2005.09318.x} {Mon. Not. R. Astron. Soc. {\bf 362}, 505 (2005).}

\bibitem{Ande} L. Anderson et al., ``The clustering of galaxies in the SDSS-III Baryon Oscillation Spectroscopic Survey: a large sample of mock galaxy catalogues,''
\doi{10.1093/mnras/sts084} {Mon. Not. R. Astron. Soc. {\bf 428}, 1036 (2013).}

\bibitem{expansao_ace} G. Goldhaber and S. Perlmutter, ``A study of 42 Type Ia supernovae and a resulting measurement of {$\Omega$}$_{M}$ and {$\Omega$}$_\Lambda$'',
\doi{10.1016/S0370-1573(98)00091-X} {Phys. Rep. {\bf 307}, 25 (1998).}

\bibitem{expansao_ace2} S. Perlmutter, et. al., ``Measurements of Omega and Lambda from 42 high redshift supernovae'',
\doi{10.1086/307221} {Astrophys.J. {\bf 517},:565 (1999).}

\bibitem{faraoni1} T. P. Sotiriou and V. Faraoni, ``$f(R)$ theories of gravity'', 
\doi {10.1103/RevModPhys.82.451} {Rev. Mod. Phys. {\bf 82}, 451 (2010).}

\bibitem{newtonian_limit} T. P. Sotiriou, ``Unification of inflation and cosmic acceleration in the palatini formalism,''
\doi{10.1103/PhysRevD.73.063515} {Phys. Rev. D \bf{73}, 063515 (2006).}

\bibitem{gravitational_stability} V. Faraoni, ``Matter instability in modified gravity,''
\doi{10.1103/PhysRevD.74.104017} {Phys. Rev. D {\bf 74}, 104017 (2006).}

\bibitem{solar_sistem} W. Hu and I. Sawicki, ``Models of f(R) Cosmic Acceleration that Evade Solar-System Tests,''
\doi{10.1103/PhysRevD.76.064004} {Phys. Rev. D {\bf 76}, 064004, (2007).}

\bibitem{inflation_FR} S. Nojiri and S. D. Odintsov, ``Modified f(R) gravity unifying R**m inflation with Lambda CDM epoch,''
\doi{10.1103/PhysRevD.77.026007} {Phys. Rev. D {\bf 77}, 026007 (2008).}

\bibitem{PhysRevD.84.024020} T. Harko, F. S. N. Lobo, S. Nojiri, and S. D. Odintsov, ``$f(R,T)$ gravity,'' 
\doi {10.1103/PhysRevD.84.024020} {Phys. Rev. D {\bf 84}, 024020 (2011).}

\bibitem{quantum_cosmology} S. Jalalzadeh, S. M. M. Rasouli, and P. V. Moniz, ``Quantum cosmology, minimal length, and holography,''
\doi{10.1103/PhysRevD.90.023541} {Phys. Rev. D {\bf  90}, 023541 (2014).}

\bibitem{energy_conditions} F. G. Alvarenga, M. J. S. Houndjo, A. V. Monwanou, and J. B. C. Orou, ``Testing some f(R,T) gravity
models from energy conditions,''
\doi{10.4236/jmp.2013.41019} {J. Mod. Phys. {\bf  4}, 130 (2013).}

\bibitem{palatini_motivation1} G. J. Olmo, ``Palatini approach to modified gravity: $f(R)$ theories and beyond,''
\doi{10.1016/j.physletb.2007.01.003} {Int. J. Mod. Phys. D {\bf  20}, 413 (2011).}

\bibitem{fR_palatini_motivation2} G. J. Olmo and H. Sanchis-Alepuz, ``Hamiltonian formulation of palatini $f(R)$ theories \`a la Brans-Dicke theory,'' 
\doi{10.1103/physrevd.83.104036} {Phys. Rev. D {\bf 83}, 104036  (2011).}

\bibitem{primeiro_frt} T. Harko, T. Koivisto, F. Lobo, and G. Olmo, ``Metric-palatini gravity unifying local constraints and late-time cosmic acceleration,''
\doi{10.1103/PhysRevD.85.084016} {Phys. Rev. D {\bf  85}, 084016 (2011).}

\bibitem{frtRT_motivation1} Z. Yousaf, M. Zaeem-ul Haq Bhatti, and U. Farwa, ``Stability analysis of stellar radiating filaments,''
\doi{10.1088/1361-6382/aa73b9} {Class. Quant. Grav. {\bf 34}, 145002 (2017).}

\bibitem{frtRT_motivation2} Z. Yousaf, M. Z. ul Haq Bhatti, and U. Farwa, ``Role of $f(R,T,R_{\mu \nu} T^{\mu \nu})$ model on the stability of cylindrical stellar model,''
 \doi{10.1140/epjc/s10052-017-4923-5} {Eur. Phys. J. C {\bf  77}, 359 (2017).}
 
\bibitem{frtRT_motivation3} M. Sharif and A. Waseem, ``On the stability of Einstein universe in $f(R,T,R_{\mu \nu} T^{\mu \nu})$ gravity,''
\doi{10.1142/S0217732318502164} {Mod. Phys. Lett. A {\bf 33}, 1850216 (2018).}

\bibitem{frtRT_motivation4} S. D. Odintsov and D. Saez-Gomez, ``$f(R,T,R_{\mu \nu}T^{\mu \nu})$ gravity phenomenology and CDM universe,'' 
\doi{10.1016/j.physletb.2013.07.026} {Phys. Lett. B {\bf 725}, 437 (2013).}

\bibitem{frtRT_motivation5} M. Zubair and M. Zeeshan, ``Cosmic evolution in the background of non-minimal coupling in $f(R,T,R_{\mu \nu} T^{\mu \nu})$ gravity''
\doi{10.1007/s10509-018-3471-2} {Astrophys Space Sci {\bf 363}, 248 (2018).}

\bibitem{godel} K. G\"{o}del, ``An example of a new type of cosmological solutions of Einstein's field equations of gravitation,''
\doi{10.1007/BF00759840} {Rev. Mod. Phys. {\bf 21}, 447 (1949).}

\bibitem{tipo_godel1} M. J. Rebou\c{c}as and J. Tiomno, ``Homogeneity of Riemannian space-times of G\"{o}del-type,'' 
\doi{10.1103/PhysRevD.28.1251} {Phys. Rev. D {\bf 28}, 1251 (1983).}

\bibitem{ctc1} J. R. Gott, ``Closed timelike curves produced by pairs of moving cosmic strings: Exact solutions,''
\doi{10.1103/PhysRevLett.66.1126} {Phys. Rev. Lett. {\bf 66}, 1126 (1991).}

\bibitem{ctc2} R. P. Kerr, ``Gravitational field of a spinning mass as an example of algebraically special metrics,'' 
\doi{10.1103/PhysRevLett.11.237} {Phys. Rev. Lett. {\bf 11}, 237 (1963).}

\bibitem{fr_and_godel} M. J. Rebou\c{c}as and J. Santos, ``G\"{o}del-type universes in $f(R)$ gravity,''
\doi{10.1103/physrevd.80.063009} {Phys. Rev. D{\bf  80}, 063009 (2009).}

\bibitem{kessence_and_godel} J. G. da Silva and A. F. Santos,``G\"{o}del and G\"{o}del-type universes in k-essence theory,'' 
\doi{10.1140/epjp/s13360-019-00065-4} {Eur. Phys. J. Plus {\bf 135}, 22 (2020).}

\bibitem{chersimon_and_godel1} C. Furtado, T. Mariz, J. R. Nascimento, A. Y. Petrov, and A. F. Santos,``G\"{o}del solution in modified gravity,'' 
\doi{10.1103/PhysRevD.79.124039} {Phys. Rev. D {\bf 79}, 124039 (2009).}

\bibitem{chersimon_and_godel2} C. Furtado, J. Nascimento, A. Petrov, and A. Santos, ``Dynamical Chern-Simons modified gravity, G\"{o}del
universe and variable cosmological constant,'' 
\doi{10.1016/j.physletb.2010.09.002} {Phys. Lett. B {\bf 693},  494 (2010).}

\bibitem{ft_and_godel} G. Otalora and M. J. Rebou\c{c}as, ``Violation of causality in $f(T)$ gravity,''
\doi{10.1140/epjc/s10052-017-5367-7}  {Eur. Phys. J. C {\bf 77}, 799 (2017).}

\bibitem{frt_and_godel} A. Santos and C. Ferst, ``G\"{o}del-type solution in $f(R,T)$ modified gravity,''
\doi{10.1142/S0217732315502144} {Mod. Phys. Lett. A {\bf 30}, 1550214, (2015).}

\bibitem{bumblebeee_and_godel} A. F. Santos, W. D. R. Jesus, J. R. Nascimento and A. Yu. Petrov. ``G\"{o}del solution in the bumblebee gravity,''
\doi{10.1142/S021773231550011X} {Mod. Phys. Lett. A {\bf 30}, 1550011 (2015).}

\bibitem{horava_and_godel} J. Fonseca-Neto, A. Petrov, and M. Rebou\c{c}as, ``G\"{o}del-type universes and chronology protection in
Horava-Lifshitz gravity,'' 
\doi{10.1016/j.physletb.2013.07.018} {Phys. Lett. B {\bf 725}, 412 (2013).}

\bibitem{brans_and_godel} J. Agudelo, J. Nascimento, A. Petrov, P. Porf\'irio, and A. Santos, ``G\"{o}del and G\"{o}del-type universes in
Brans-Dicke theory,'' 
\doi{10.1016/j.physletb.2016.09.011} {Phys. Lett. B {\bf 762}, 96 (2016).}

\bibitem{frq_and_godel} F. Gama, J. Nascimento, A. Petrov, P. Porf\'irio, and A. Santos, ``G\"{o}del-type solutions within the $f(R,Q)$
gravity,''
\doi{10.1103/physrevd.96.064020} {Phys. Rev. D {\bf 96}, 064020 (2017).}

\bibitem{frt_palatini} J. Gon\c{c}alves and A. Santos, ``G\"{o}del and G\"{o}del-type solutions in the Palatini $f(R,T)$ gravity theory,'' 
\doi{10.1142/S0218271821500140} {Int. J. Mod. Phys. D {\bf 30}, 2150014 (2021).}

\bibitem{fr_phi_X} J. Gon\c{c}alves and A. Santos, ``A study on causality in $(R,\phi,X)$ theory,''
\doi{10.1142/S0217751X21500093} {Int. J. Mod. Phys. A {\bf 36}, 2150009 (2021).}

\bibitem{reviewbook} T. Harko and F. S. N. Lobo,  \textit{Extensions of $f(R)$ Gravity - 
Curvature-Matter Couplings and Hybrid}, 1st ed. (Cambridge University Press, UK, 2019).

\bibitem{godel_fRT} A. F. Santos, ``G\"{o}del solution in $f(R,T)$ gravity,''
\doi{10.1142/S0217732313501411} {Mod. Phys. Lett. A {\bf 28}, 1350141, (2013).}

\bibitem{tipo_godel_frt} A. F. Santos and C. J. Ferst, ``G\"{o}del-type solution in $f(R,T)$ modified gravity,''
\doi{10.1142/S0217732315502144} {Mod. Phys. Lett. A {\bf 30}, 1550214 (2015).}

\bibitem{fr_hybrid} J. Santos, M. J. Rebou\c{c}as, and A. F. F. Teixeira, ``Homogeneous G\"{o}del-type solutions in hybrid metric-
Palatini gravity,''
\doi{10.1140/epjc/s10052-018-6025-4} {Eur. Phys. J. C {\bf 78}, 567 (2018).}

    
\end{thebibliography}
\end{document}